# Codeshare agreements between airlines: literature review with the aid of artificial intelligence

Lucas T. B. Mendes
Alessandro V. M. Oliveira✈
Aeronautics Institute of Technology, São José dos Campos, Brazil
✈ Corresponding author. Praça Marechal Eduardo Gomes, 50. 12.280-250 - São José dos Campos, SP - Brazil.
Email address: lucasltbm@ita.br.

***Abstract*: Codeshare agreements are contracts that allow two or more airlines to share seats on the same flight. These agreements, which are widespread in commercial aviation as a response to highly competitive environments, have enabled the expansion of airline networks without additional costs or risks for the companies involved. The literature presents ambiguous effects associated with the practice, with evidence of increased supply and reduced prices in situations of route complementarity, while also pointing to anti-competitive impacts in markets where companies act as competitors. A review of scientific production over time, including theoretical contributions and case studies, is essential to understand the evolution of these agreements and their implications, especially in the Brazilian context, marked by its own characteristics and particular regulatory history. Thus, this article reviews the literature on codesharing, with an emphasis on the Brazilian market, and uses the Litmaps computational tool, based on artificial intelligence techniques, to support the contextual analysis of publications through their citation relationships. The ultimate goal is to identify and evaluate the main evidence accumulated over decades on the effects of these agreements in Brazil. The joint analysis of the contributions allows us to outline the current state of knowledge, characterize specificities observed in the Brazilian market, and identify gaps that may guide future studies.**

## I. INTRODUCTION

For decades, codeshare agreements have played a central role in the global airline network, illustrating how companies from different countries and alliances expand their connectivity through commercial cooperation. These arrangements remain relevant because they allow companies to expand their presence in foreign markets without having to operate all routes directly, a practice that has become even more strategic in the face of growing international competition. By definition, codeshare agreements are cooperation contracts between two or more airlines, in which one or more of them can sell seats as a "marketing airline" on flights operated by another company, the "operating airline" (Alderighi et al., 2015). In this way, companies began to market and obtain revenue on routes operated by third parties, expanding their service networks and reducing logistical and regulatory barriers, especially after the Deregulation Act of 1978, a milestone in the deregulation of the American market.

This cooperation mechanism became consolidated in both the American and European markets, driven by the spread of agreements on numerous international routes. One of the central factors in this expansion was the reduction of regulatory barriers so that airlines could market their services in foreign markets through cooperation agreements, as highlighted by Brueckner (2001). With the advancement of this practice, different types of codesharing began to be adopted in various countries, each adapted to the operational and regulatory characteristics of each market. The possibility of expanding the network of routes served without assuming additional costs or risks and, at the same time, improving indicators such as flight utilization factors, became an important incentive for the growing adherence to these arrangements.

The evidence in the literature shows that studies on codesharing present mixed and ambiguous results, in the sense that some analyses point to possible benefits to the market and society, while others suggest potentially negative effects. The positive effects discussed in theoretical studies are usually associated with reduced costs and risks for airlines, gains in operational scale, and the consequent possibility of expanding supply and reducing prices, especially on routes with lower commercial attractiveness. On the other hand, there is a consistent set of studies that identifies different results when codesharing is applied on routes where companies act as competitors. In these cases, the observed effects tend to reflect the anti-competitive nature of cooperation, generating possible increases in market power and reducing competitive pressure between the companies involved.

In this context, a review of the literature on codesharing, considering theoretical contributions and their applications over the years, is important to allow a comprehensive understanding of the topic and its interfaces, highlighted due to the typical agglomeration of codesharing practices to more comprehensive strategic alliance agreements between companies, which tends to result in inaccurate and non-individualized conclusions (Sampaio and Urdanoz, 2022).

In the Brazilian context, the adoption of codesharing as a market strategy had its own particularities; the Brazilian civil aviation industry is one of the few examples where there was a brief period of market reregulation after a few years of deregulation. As highlighted by Bettini and Oliveira (2008), to combat an alleged systemic excess of capacity and exacerbated competition in the airline market, the Brazilian Federal Government, starting in 2003, took a stance aimed at increasing market regulation, in addition to encouraging the signing of a codeshare agreement between two of the largest Brazilian airlines: TAM and Varig.

Since the first codeshare agreements signed in Brazil in the late 1990s, several examples have been signed and implemented in the Brazilian market, in different contexts, scopes of cooperation, cooperative routes, and partnership periods. Considering the diverse and ambiguous effects of these agreements on society, in terms of supply and price, and the importance of conducting and analyzing empirical studies on these devices, this study focuses on evaluating recent scientific contributions regarding the analysis of the effects of these agreements on the Brazilian market.

For a proper review of the literature on any topic or field of scientific knowledge, the use of diverse tools and methodologies is important in order to properly evaluate the research base as a whole, its characteristics, and perspectives on the topic. Furthermore, the development in recent years of artificial intelligence computational tools has extended to use in various educational and research fields to enhance the understanding and analysis of scientific data and learning and teaching processes (Chen et al., 2020). There is ample evidence of the positive impacts associated with the use of computational tools in various institutions, including research and teaching projects (Mabic et al., 2022).

Thus, the main objective of this study is to conduct a literature review on codeshare agreements, focusing on identifying and analyzing empirical articles on the recent Brazilian context, in order to gather the main evidence on the effects of these practices on society. It is important that the diversity of the effects of these agreements, widely discussed in the literature, be investigated in the context of their application to Brazilian routes or airlines.

The Litmaps computational tool will be used as support to optimize graphic visualization and search for scientific articles for the Review Databases. Litmaps is a web platform that provides a graphic representation based on literature review, allowing users to identify themes and perspectives, as well as visualize and explore the relationships between different concepts and ideas within a given field. Through natural language processing algorithms and machine learning, the application analyzes the text of academic articles and creates a visual representation of the literature review (Sulisworo, 2023). In addition to organizing citations and connections between publications, Litmaps uses artificial intelligence techniques to estimate thematic similarities, suggest relevant articles, and recognize patterns within large databases, which expands the capacity to explore the literature and facilitates understanding of its structure.

The structure of the article is as follows. After this introduction, section II describes the methodology, detailing the procedures for searching, selecting, and analyzing the studies, as well as the tools used. Section III, dedicated to the literature on codeshare agreements, discusses the main works identified, including the initial review base, the final study base, and the studies applied to the Brazilian context. Section IV examines the patterns observed in the Brazilian literature on codeshare agreements, highlighting recurring characteristics and limitations. Finally, section V brings together the conclusions, summarizing the findings, identifying gaps, and suggesting avenues for future research.

# II. Methodology

The literature on codeshare agreements in air transport is extensive and well-established, with a large volume of studies ranging from theoretical foundations to empirical analyses in different markets. As in many areas of knowledge, the continuous expansion of this scientific output makes it increasingly necessary to use artificial intelligence tools to organize information, compile results, and identify relevant patterns within extensive databases. In this context, the research and collection of articles that comprise this review were directed at studies that explicitly address codesharing or that apply methodologies in contexts in which the practice of codesharing or flight sharing is present, both in international scenarios and in the Brazilian market. Publications in recognized journals and forums were consulted, obtained through searches in the CAPES Journal, Elsevier, Google Scholar, and Web of Science. Once gathered, these articles were incorporated into Litmaps to allow for a structured visualization of the database, clarifying how these works relate to each other and their relevance within the analyzed theme.

Litmaps is a digital platform for mapping and visualizing scientific literature, available at www.litmaps.com. The tool organizes articles into graphical representations that show the evolution of a field of research over time. Each publication is displayed as a circle whose position reflects its publication date and degree of relevance to the topic, while the size of the circle is proportional to the number of citations received. The lines connecting different circles represent cross-references between works. This structure allows us to clearly identify which studies are central to the literature, which authors play relevant roles in shaping the field, and how ideas spread between publications over the years. In addition, Litmaps uses natural language processing and machine learning techniques, essential components of artificial intelligence systems, to analyze article abstracts, recognize patterns, estimate thematic similarities, and suggest relevant connections or groupings. This AI feature significantly expands the database's exploration capabilities, allowing researchers to identify trends, gaps, and conceptual clusters more quickly and efficiently.

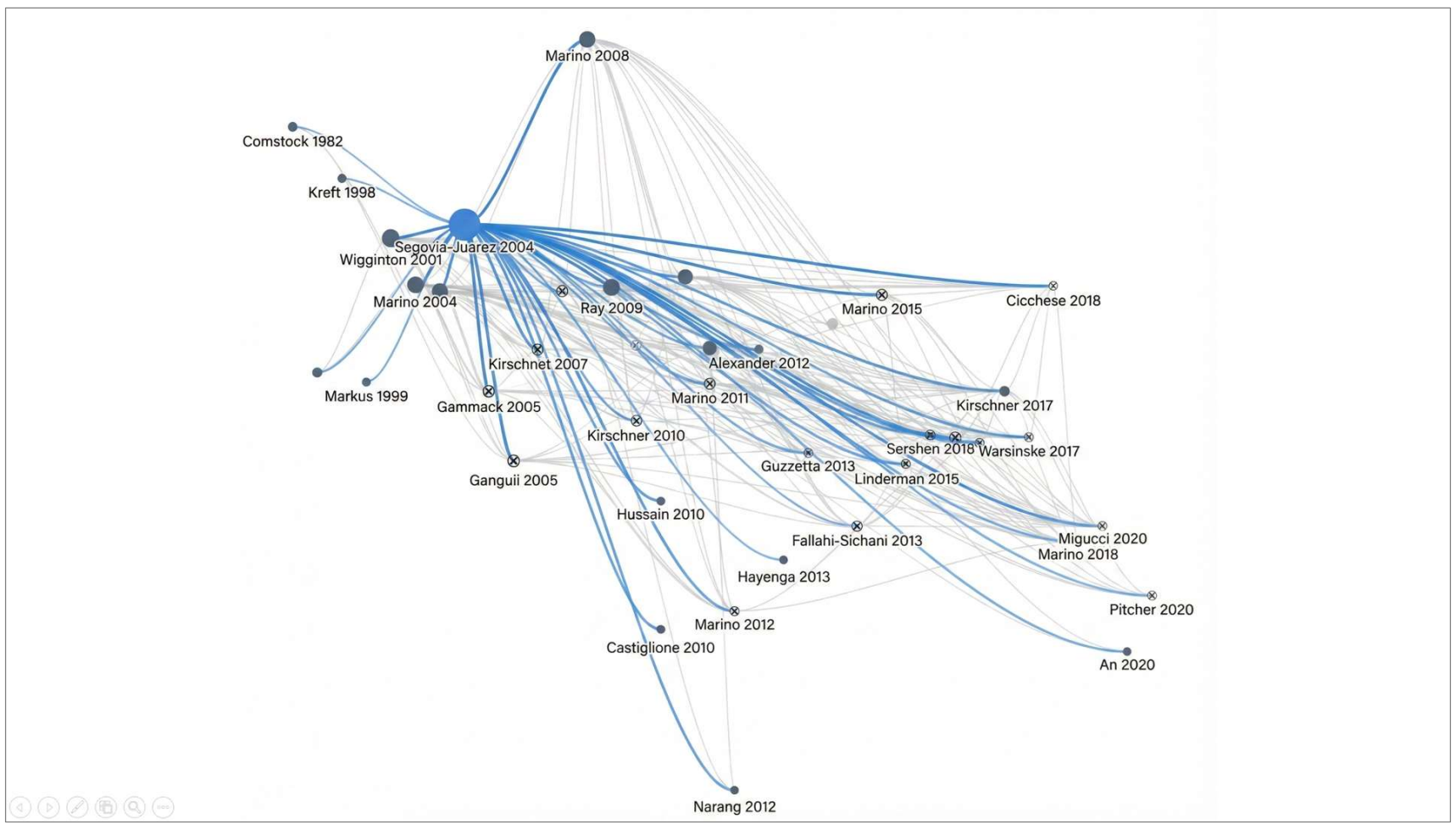

**Figure 1 - Illustration of an article's relevance map, via Litmaps (Segovia and Baumgartner, 2024)**

The graphical representations developed by the Litmaps application take into account the citation relationship between specific authors and articles; the maps generated allow researchers to visualize the evolution of research on different topics over time (Sulisworo, 2023). In Litmaps, article information is analyzed in an integrated manner, and the articles evaluated are arranged graphically according to the connectivity and importance of each publication within the topic studied. These articles are positioned according to their publication date on the X-axis and their relevance to the theme on the Y-axis. Each publication added to the database is represented by a circle, whose size is proportional to the number of citations received, while the lines connecting the circles represent cross-references between articles. Figure 1 shows an illustration generated by Litmaps that exemplifies how this works. In it, each circle corresponds to an article, and its size indicates the volume of accumulated citations. The circles positioned higher on the vertical axis represent publications considered most relevant to the topic, while the circles on the right indicate more recent studies. The lines connecting the points function as a map of references, showing which articles cite which others and revealing thematic clusters and trajectories of influence within the field. Darker or thicker lines usually indicate stronger or more frequent connections, while lighter lines suggest secondary relationships. This visualization allows the reader t ly and quickly identify central articles, research clusters, temporal evolution, and citation patterns that structure the development of the literature.

Similar to Segovia and Baumgartner (2024), the Litmaps tool will be used in this article as a tool to support the structuring of the evaluated literature review base throughout the stages of the methodology, providing an overview of the citation network and identifying the main articles and authors on the topic. This integrated visualization of articles and authors, based on insights obtained using Litmaps, will be used to further the literature review on the topic.

It is important to highlight that the use of computational tools imposes some significant limitations and assumptions on the study, which were taken into account in conducting this review. The main aspect is the search and integration of articles into publication databases, which only considers articles published in relevant, high-quality international journals, not taking into account scientific contributions to the topic from seminar articles, smaller or regional journals, doctoral theses, master's dissertations, among others. In addition, beyond the relationship traced by the tool between citations and references between authors and articles, Litmaps also has relationship and search mechanisms based on information from the content of the articles, such as theme, scope, or methodology; however, the tool is limited to evaluating the abstracts of the articles, resulting in inaccuracies in the classification due to the lack of depth in the content of the articles in their entirety. Furthermore, as the platform displays only the name of the first author followed by the year of publication on the maps by default, there is a risk of inducing a bias of simplification of citations, in which the researcher tends to mention only the main author and fail to include the other names or to correctly use the clause “et al.,” reducing the accuracy and completeness of the references.

Thus, the first step of the proposed methodology consisted of structuring the initial database of relevant articles on the topic in Litmaps. Next, an in-depth analysis of this initial database was carried out, using categorization and optimized search tools and identifying the main publications on codeshare agreements in the Brazilian context. This second stage of the methodology, analyzing the database and filtering articles of interest on the topic, resulted in the consolidation of the definitive literature review database proposed by the study.

Another relevant aspect of the literature is that codeshare agreements are typically referenced and analyzed in case studies and agreement scenarios in which their use is associated with strategic and more comprehensive alliances between airlines; the concepts evaluated by the articles must be well defined so that the review considers only aspects of codesharing practices themselves. Thus, articles that analyze cooperation agreements within the scope of strategic alliances, without distinguishing codesharing from other market practices, were disregarded from the final consolidated database.

Based on the final structured review database, the articles most relevant to the proposed study were evaluated and identified, considering the focus of empirical articles on agreements in the Brazilian market. These selected articles will therefore be the subject of in-depth analysis to gather evidence and considerations from the literature. This third and final stage will allow for an analysis of the panorama of the application of these cooperation tools in Brazil, the main objective of the study, considering the number of publications evidenced, types of agreements analyzed, and scopes of empirical study applied in the Brazilian context.

# III. LITERATURE ON CODESHARE AGREEMENTS

The US civil aviation market, historically characterized as one of the most developed and relevant in the world, is a benchmark for regulatory practices and relationships between companies in the sector. As an important milestone in the development of the airline industry in the United States, the Deregulation Act of 1978 was responsible for a disruption in the market, with the removal of federal control over fares, routes, and the entry of new companies into the market (Smith Jr. and Cox, 2025).

One of the main consequences of the Deregulation Act was increased pressure for competitiveness in the market, due to the reduction of barriers to entry and regulatory and market criteria to be met. In response to this growing competitiveness, most American companies with hub-and-spoke networks joined some form of cooperative alliance, one of these measures being codeshare agreements on routes in their network, a practice that has consolidated and expanded over the years (Ito and Lee, 2007; Yimga, 2018).

Sampaio and Urdanoz (2022) discuss in their study what they consider to be the five main reasons why airlines enter into codeshare agreements, summarizing the main aspects cited in the literature:

- It allows for the rapid expansion of the network of routes served by airlines, without additional costs or risks, even allowing access to potentially restricted markets;
- It allows cooperation for the coordination of frequencies and schedules of scheduled flights on routes where companies operate in parallel;
- It allows for the sharing of platforms, investments, and brand power of partner airlines, boosting the marketing and sales of cooperative products;
- It allows for the reduction of marginal costs through cooperation in airline route operations, such as in advertising campaigns and product promotion materials; and
- They can generate more comprehensive cooperation between airline actions and operations, such as optimized baggage and passenger processing procedures for connecting flights.

Among the various characteristics and expected effects of adopting codeshare agreements, there is a predominance of evidence from studies that consider that the positive effects on society, such as increased supply on less attractive routes and reduced airfares, would result from the reduction of marginal costs and risks for airlines, due to reduced competitiveness and uncertainties in operating on these routes. In an important study by Brueckner (2001) on the subject, the author analyzes codeshare practices in the market at the time and concludes that they can be efficient and beneficial to society, as cooperation contributes to the elimination of double marginalization. Studies such as Brueckner and Whalen (2000), Brueckner (2003), and Whalen (2007), which are highly relevant in the literature on the subject, are examples among several studies that show that prices of non-cooperative products tend to be impacted by double marginalization, while products resulting from cooperation agreements or alliances between airlines would be significantly lower.

However, the resulting impacts on markets and, above all, on society are not only positive; there is an ambiguous nature to the effects of codesharing on society in terms of supply and consumer prices, where the balance of positive and negative effects depends on the market and the format of the agreement signed (Alderighi, 2015). On point-to-point routes in networks where there is complementarity between the routes of the cooperating airlines, the reduction in marginal costs tends to lower airfares, whereas on routes where both airlines operate in parallel, the reduction in the effects of competition and greater barriers to entry for new airlines tend to have a negative impact on their fares (Goetz and Shapiro, 2012; Sampaio and Urdanoz, 2022).

The effects become even more complex when considering the various characteristics involved in the adoption of codesharing; for example, whether shared seat management is based on real-time systems or predefined balances, whether the agreements result in online, virtual, or traditional products, according to the type of shared routes and how they are operated (Brueckner, 2001; Alderighi, 2015; Yimga, 2018). The literature on the subject converges on the understanding that each agreement is signed in a specific context, considering a given regulatory structure, the positioning of airlines in the market, and operations on a given route, and influences markets in different ways.

In Brazil, the first case of cooperation between airlines occurred in 1959 and is often considered a pioneering initiative on the world stage, even preceding the consolidation of similar arrangements in corridors such as New York-Boston-Washington. Under this agreement, Varig, Cruzeiro, and Vasp began issuing a unified ticket that could be purchased from any of the companies and used on the first available flight on the Rio-São Paulo air bridge, regardless of which company operated it. This format allowed passengers to board the next departure, giving the ticket an interoperability that was unusual for the time. The agreement aimed to combat the aggressive stance of Real Aerovias in the market, an airline that allegedly engaged in predatory practices and unfair competition in the market (Folha de São Paulo, 2019). Although this case did not yet constitute a formal seat-sharing agreement on a specific route, it is generally interpreted as a preliminary operational cooperation arrangement, prior to the formats that would later be recognized as codeshare. This is because there was no structured coordination of seat supply or scheduled integration of the companies' capacities, elements characteristic of later codesharing agreements.

With regard to codesharing itself, the first practices in Brazil date back to 1998, with the agreement signed between Vasp and Transbrasil. In the early 2000s, these agreements became widespread in the Brazilian airline industry, with Brazilian airlines adopting them to overcome financial difficulties resulting from high fuel prices, among other economic factors (Lovadine and Oliveira, 2005; Britto, 2020).

Fast forwarding another twenty years in history, the Brazilian market continued to see codeshare agreements over the years: an agreement between Gol and American Airlines since 2019, allowing the operation of international routes by foreign airlines; a brief agreement signed between Azul and Latam between 2020 and 2021, contemporary to the period of major restrictions on air operations due to COVID-19; discussions about a codeshare agreement to be signed between Azul and Latam began in 2024, but face a complex scenario for establishment. The history of the dissemination of these practices in Brazil over the last few years, which are highly relevant and evident in the market, has been accompanied by the development of publications and scientific production on the subject, delimiting the theme that was taken as the final scope of the following review.

## III.1. INITIAL BASIS FOR LITERATURE REVIEW

As a starting point for the proposed study, the structuring of the initial basis for the literature review took into account the addition of the main articles and authors on the broad topic of Strategic Alliances and codeshare agreements, considering empirical and theoretical studies, under national or international route agreements, throughout the world . Due to the similarities in the conditions of application and case examples of codeshare agreements and broader alliances together, both in the market and in the literature, several publications encompass codeshare agreements within broader alliances, or develop studies on codeshare, but without distinguishing them from antitrust immunity agreements, with additional mechanisms.

The research of these files was carried out using conventional and consolidated sources of scientific articles, but was mainly based on the use of the search tool integrated into Litmaps, as the database was being built. Thus, this initial research included articles on codeshare itself, as well as relevant examples from the literature that analyze codeshare associated with alliances, in order to prioritize the consolidation of a comprehensive initial database and so that the Litmaps search tool could use the comprehensive database as a reference for various articles dealing with codeshare in structuring the initial database of articles.

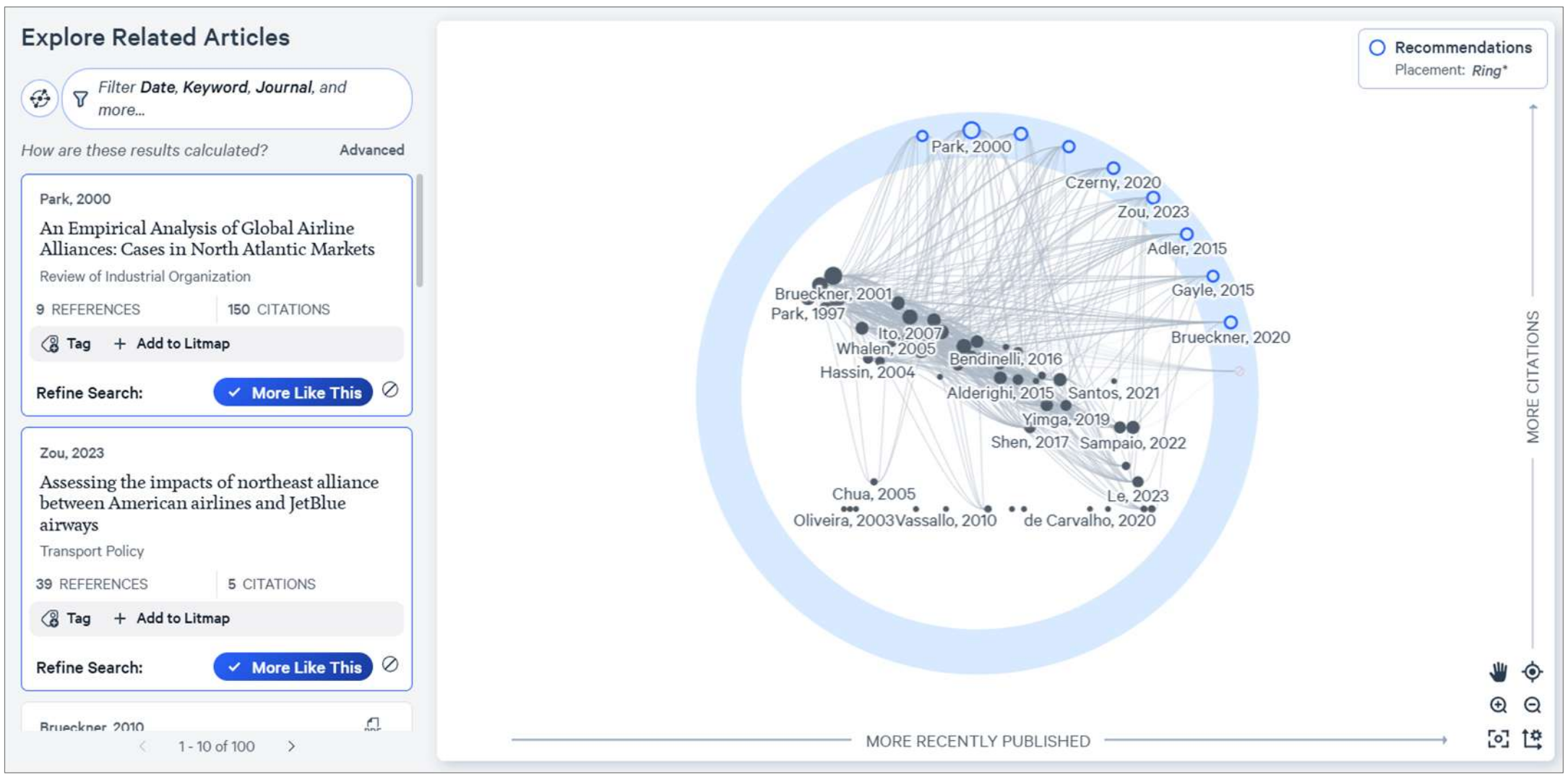

**Figure 2 - Litmaps article search tool, based on the synergy between references among the articles in the database (Own elaboration. Source: Litmaps)**

This research stage resulted in a database of 60 articles, graphically arranged in Litmaps as shown in Figure 1 below. The articles are presented in dark gray circles, with sizes proportional to their interconnectivity of citations and references within the database, and positioned according to their relevance to the topic (total number of references) on the vertical axis and date of publication on the horizontal axis. For example, "Gualini, 2024" is presented in medium size, with significant connectivity to articles in the database, but has few total references considering other publications. Thus, the circle representing this article is positioned to the right and below in the representation of the database. These criteria for the visual layout of articles in the Litmaps database will be maintained in the other analyses, images, and versions of the scientific article database to be presented in this study.

With the consolidation of the initial review database, the analysis of the added articles leads to the identification of several relevant articles on the topic. The articles found are mostly focused on the study of codeshare agreements, antitrust immunity agreements, and measures prior to large mergers or acquisitions between airlines, considering the American market and mainly international routes.

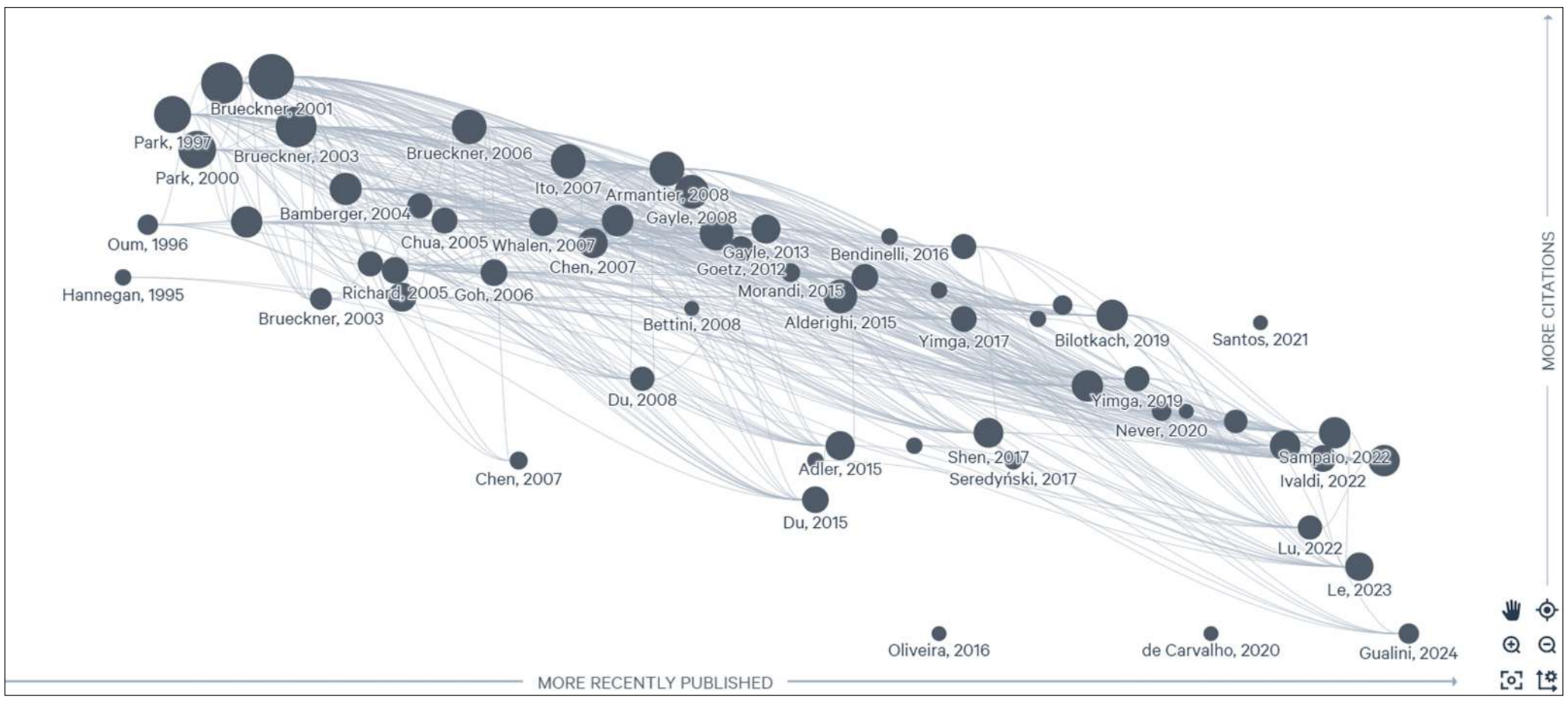

**Figure 3 - Initial database of scientific articles for the study, considering codeshare agreements and Strategic Alliances between airlines (Own elaboration. Source: Litmaps)**

Among these articles, Park and Zhang (2000) is an example of an article that was initially added to the review base but does not distinguish these agreements in its study. Although quite relevant to the topic, considering its position on the initial database map and number of references in the literature, its contribution to the assessment of consumer effects on airfares, demand, and consumer surpluses, considering four major alliances between airlines in the North Atlantic market and data analysis between 1990 and 1994, he did not highlight the practice of codesharing in his empirical study, thus not allowing for the assessment of the influence of these agreements, as proposed in the following stages of this review.

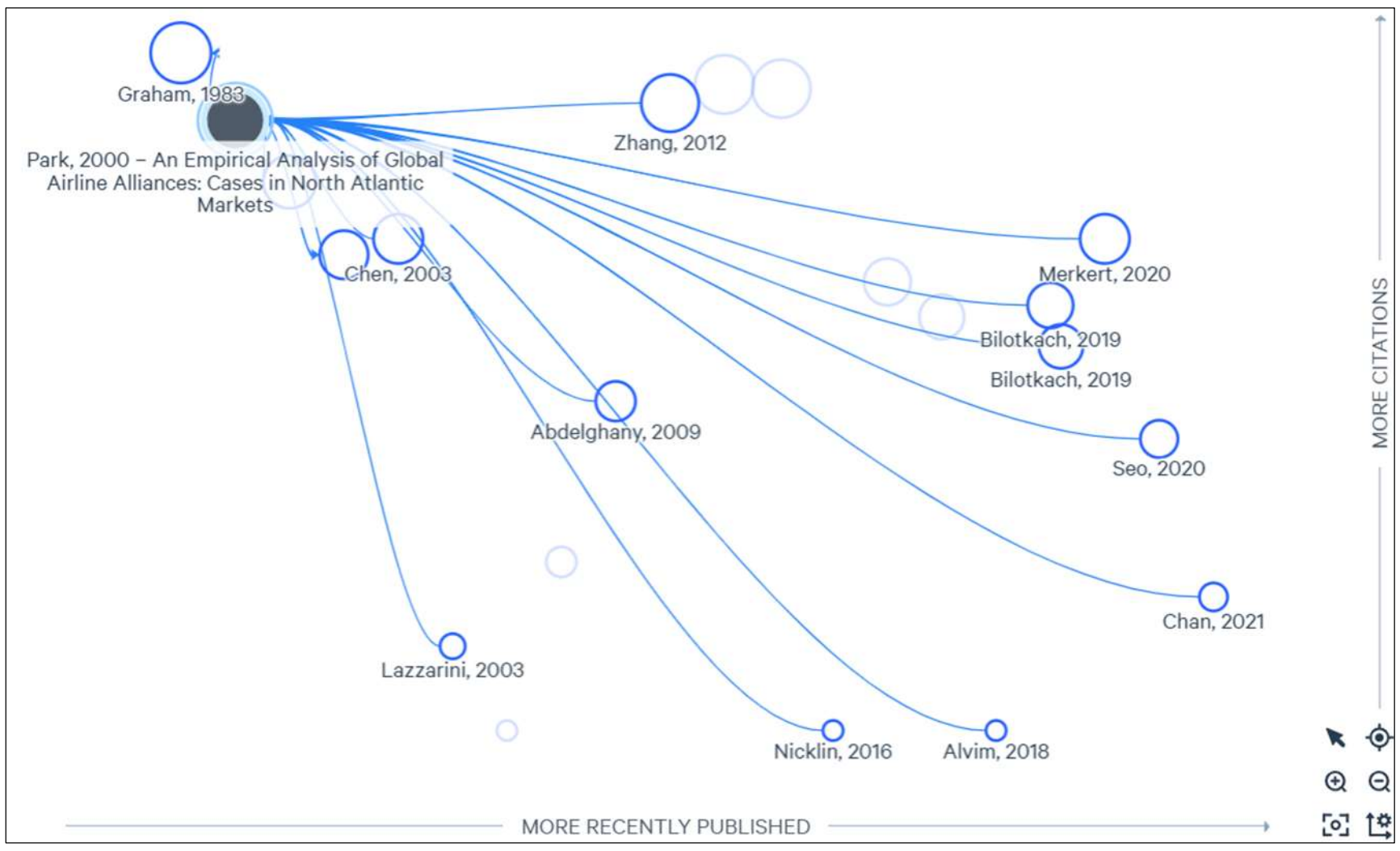

**Figure 4 - Graphical representation of Park (2000) and other articles with shared references (Own elaboration. Source: Litmaps)**

Other more recent examples of articles identified in the initial database with a high degree of relevance, but which do not distinguish the specific aspects and effects of codeshare in their studies, are: Zou et al. (2023), who assess the impacts of the alliance between American Airlines and JetBlue in terms of market concentration and airfares, with analysis of data between 2019 and 2021 from relevant airport routes in Boston and New York, and; Ivaldi et al. (2022), who use a specific methodology to analyze the effects of alliances on the distribution of airfares in the domestic market, with analysis of data from the Airline Origin and Destination Surveys (DB1B) between 2008 and 2019. Figures 5 and 6 below show the graphical representations of the articles in Litmaps, considering their connections through references to other articles and indicating their high degree of relevance to the topic. These studies, among others identified in the initial database that do not specify codeshare agreements, will be selected and excluded from the article database as the next step in the proposed methodology to consolidate the final review database.

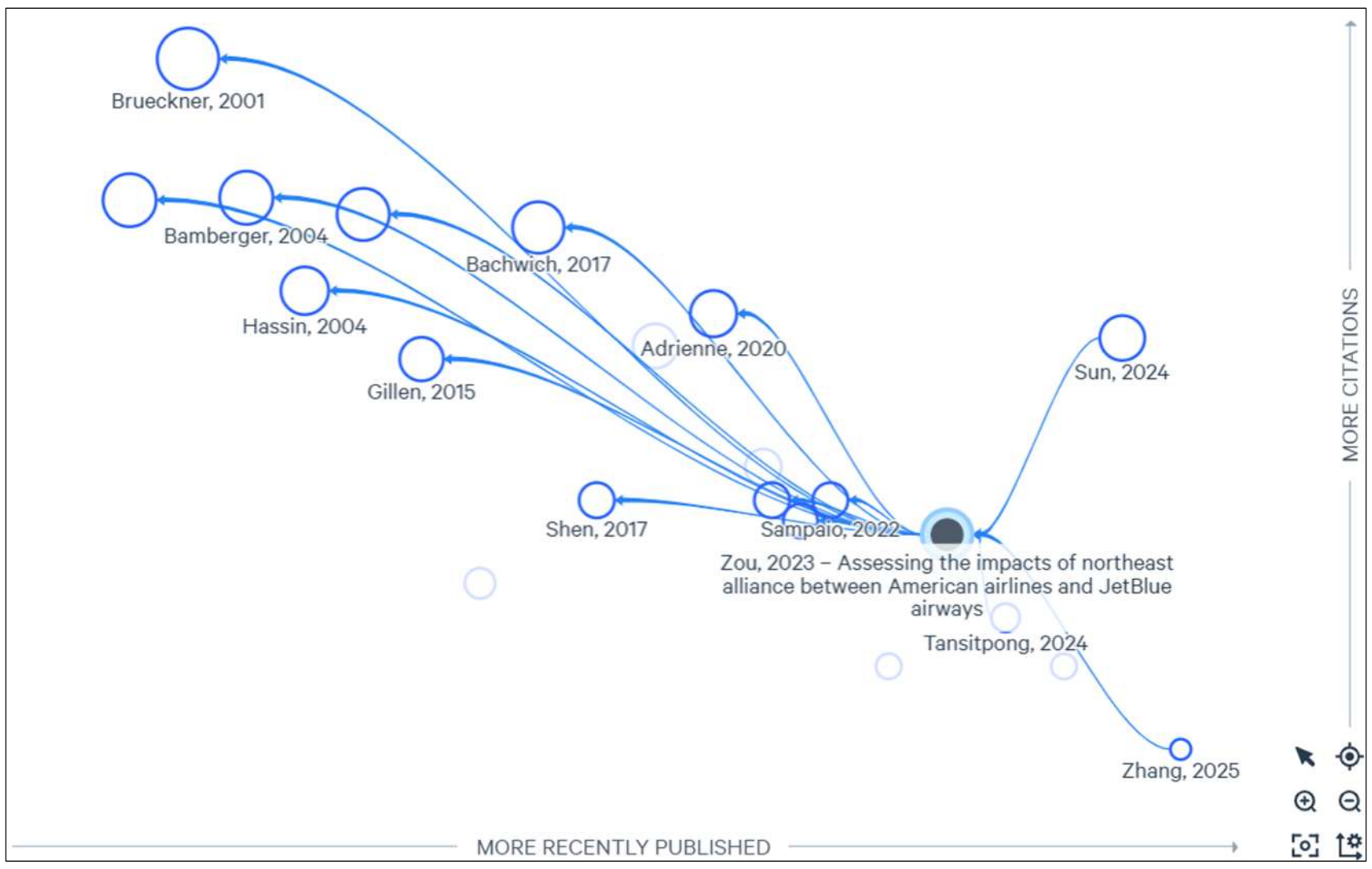

**Figure 5 - Graphical representation of the article Zou (2023) and other articles with shared references (Own elaboration. Source: Litmaps)**

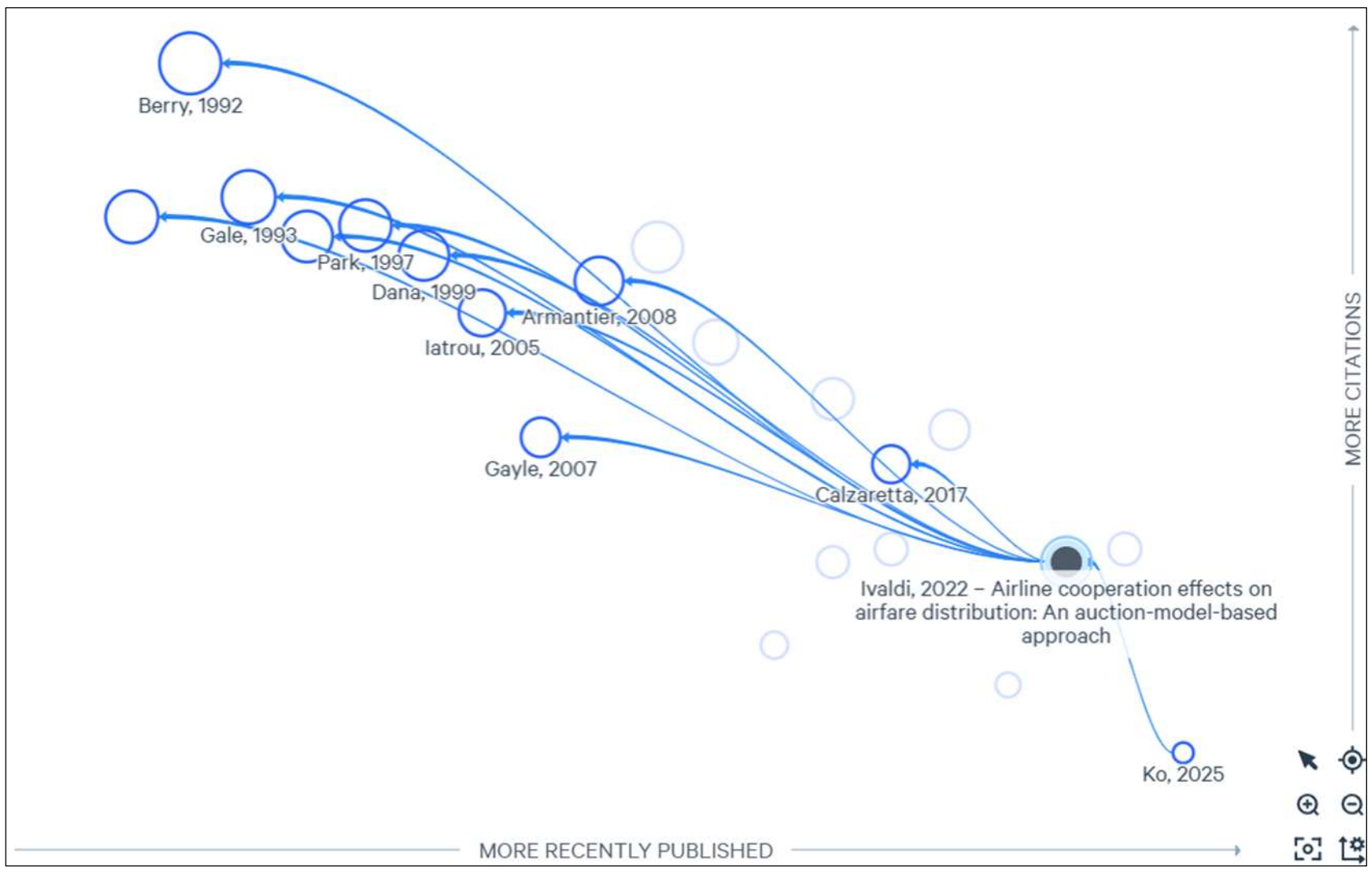

**Figure 6 - Graphical representation of the article by Ivaldi (2022) and other articles with shared references (Own elaboration. Source: Litmaps)**

## III.2. FINAL STUDY DATABASE

The next step in the proposed methodology consists of structuring the final database of articles on the topic and identifying the articles most relevant to the chosen focus. It began with the exclusion of articles that do not differentiate codeshare agreements from other alliances analyzed in their studies. Following several stages of research, carried out using the search tool integrated into Litmaps, the final database was obtained, as shown in Figure 6 below, containing 54 articles on the topic, with seven articles identified as most relevant highlighted in green, given the objective of analyzing the Brazilian context.

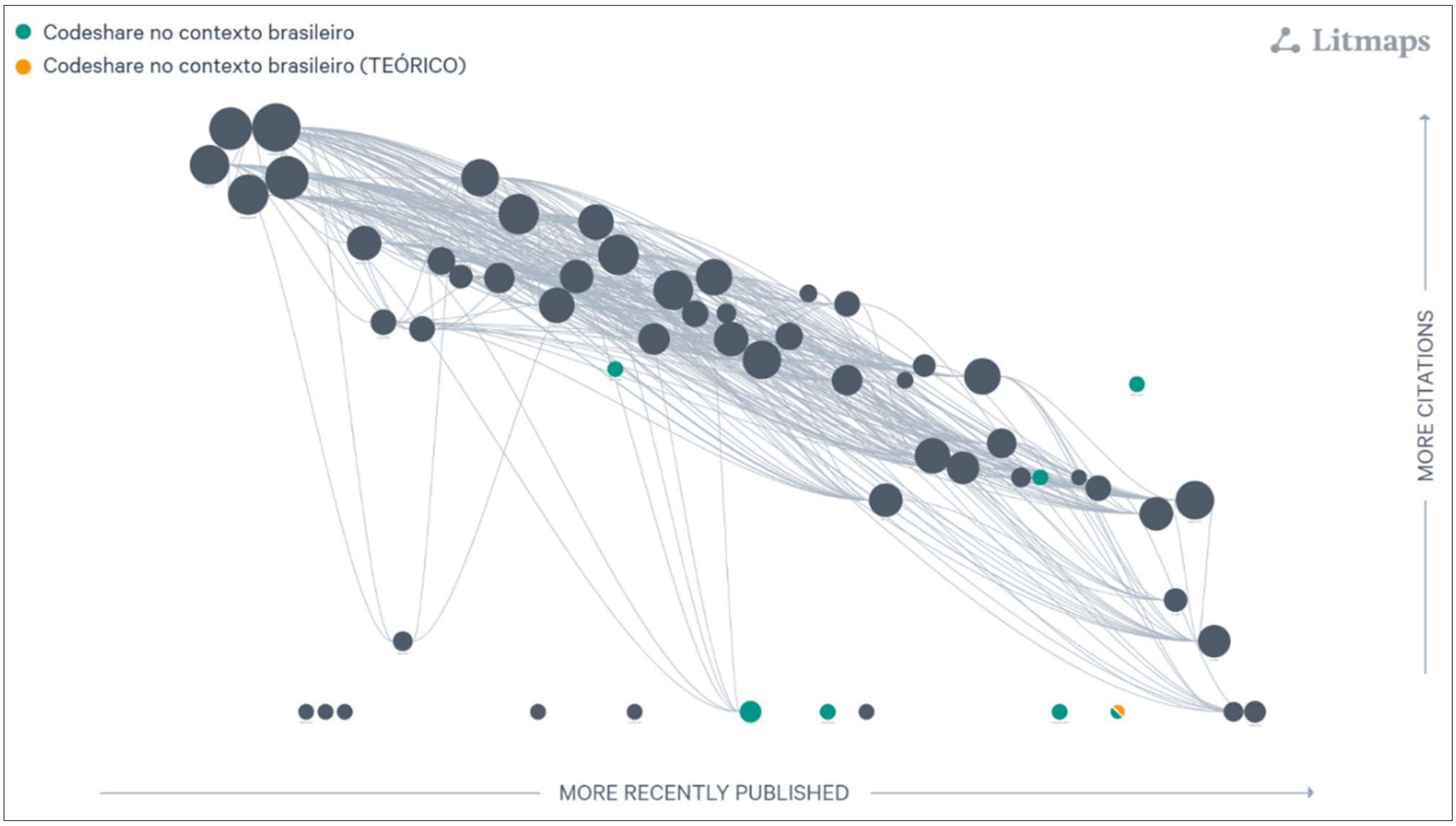

**Figure 6 - Final database of scientific articles for the study, considering codeshare agreements in particular and highlighting relevant articles identified (Own elaboration. Source: Litmaps)**

In this final database, it is important to highlight the presence of articles of high relevance to the topic, considering the categorization made by the Litmaps tool, which serve as a benchmark on the literature on codesharing and contribute to "anchoring" the intelligent research databases of the Litmaps tool in these examples among the main scientific contributions on the subject. The database contains articles published from 2000 to the present day, with contributions from important authors over the years, such as:

Jan K. Brueckner, who has made relevant theoretical contributions related to the first global codeshare agreements at the end of the 20th century (Brueckner and Whalen, 2000; Brueckner, 2001; Brueckner, 2003; Brueckner et al., 2011);

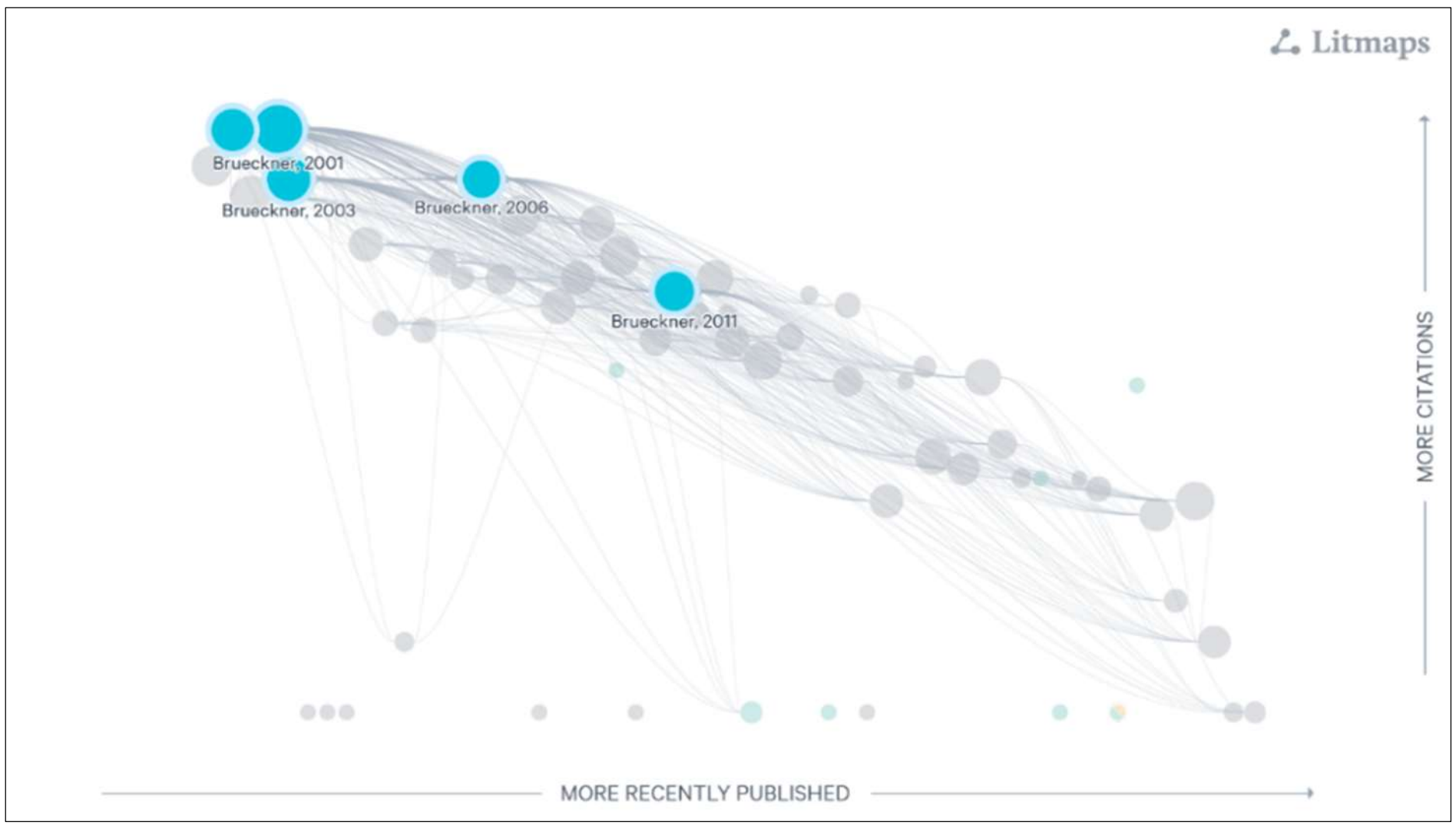

**Figure 7 - Publications by author Jan K. Brueckner highlighted among the articles in the final database (Own elaboration. Source: Litmaps)**

Philip G. Gayle, who has conducted several empirical studies on the econometric effects of codeshare agreements and broader alliances on the market and society, with a focus on the American market (Gayle, 2007; Gayle, 2008; Gayle, 2013; Gayle and Brown, 2015; Gayle and Xie, 2019)

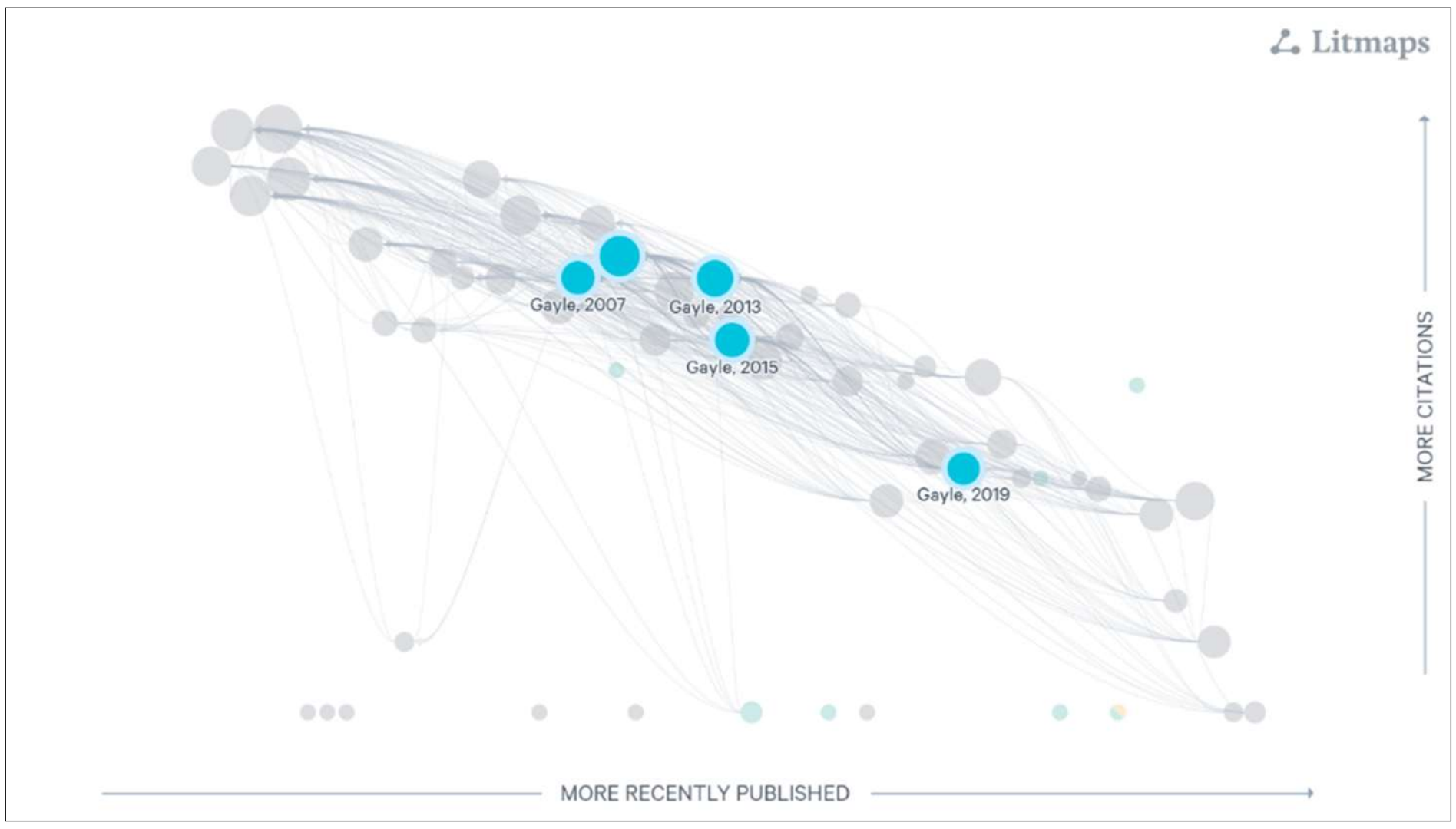

**Figure 8 - Publications by author Philip G. Gayle highlighted among the articles in the final database (Own elaboration. Source: Litmaps)**

Jules O. Yimga, who has examples of empirical articles that analyze various effects of codeshare and cooperation practices in the airline industry: in addition to econometric studies, Yimga proposes analysis of aspects such as delays and OTP performance, and quality of cooperative products, with several recent publications on the topic (Yimga, 2017; Yimga, 2018; Yimga and Gorjidooz, 2019; Yimga, 2022);

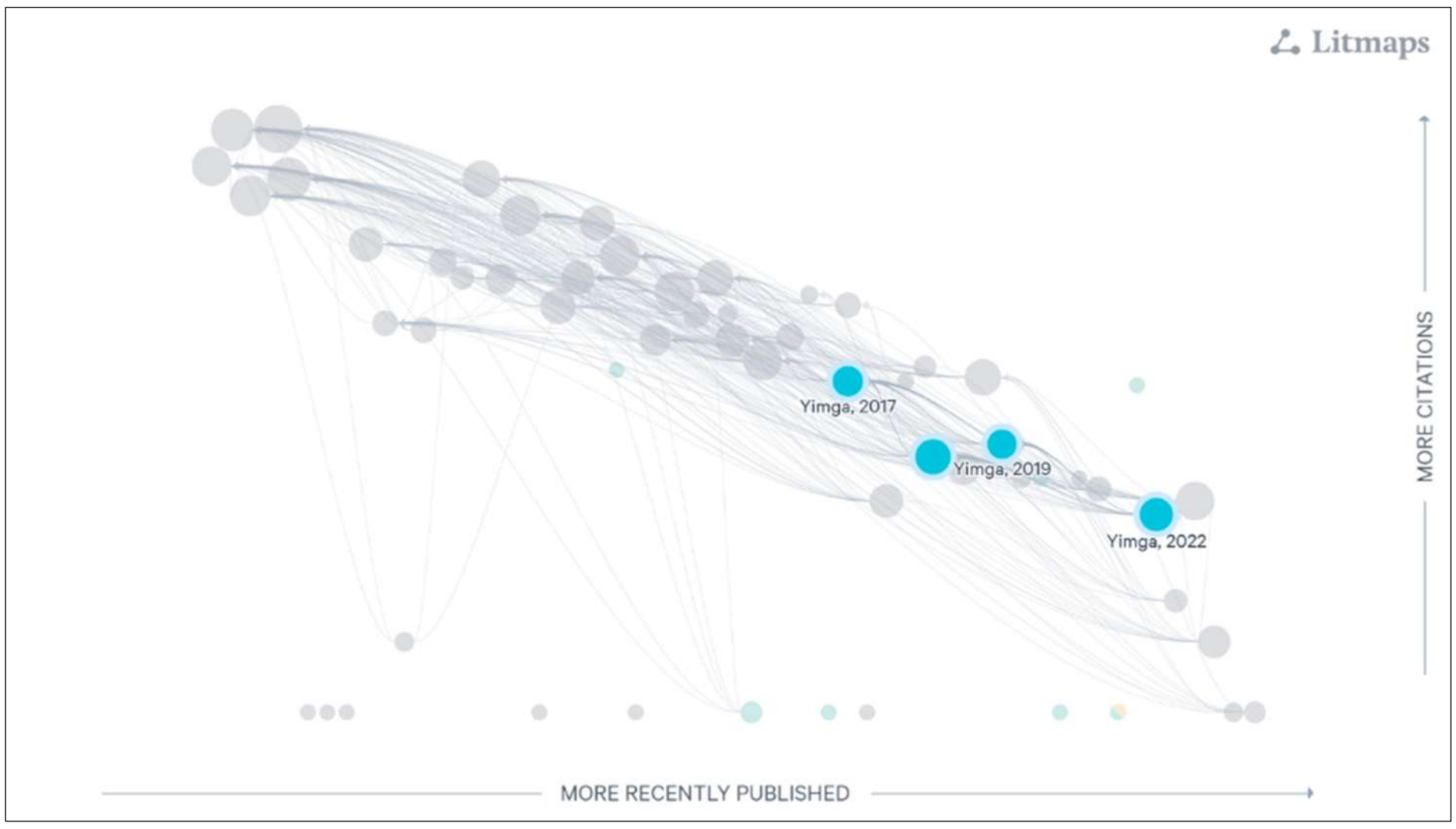

**Figure 8 - Featured publications by author Jules O. Yimga, among the articles in the final database (Own elaboration. Source: Litmaps)**

An important aspect of the methodology is the fact that the articles that comprised the final database and came from the more comprehensive initial database, when used as the intermediate research database by Litmaps for the progressive composition of the final database, did not generate significant numbers of new articles related to the Brazilian context. A direct search for articles on codeshare agreements in Brazil was conducted in Litmaps to overcome this initial difficulty in adding articles on the Brazilian context using the search tool from the article database.

Combinations of the terms "Codeshare," "Brazil," and "Airlines" were used, and a total of 10 articles related to the Brazilian airline industry were found and added to the final database, contributing with analyses of codeshare agreements and cooperation in general with regional relevance to the literature on the subject, but which have a low degree of relevance according to Litmaps. These articles were important in adding Brazilian examples and the subsequent selection of the main articles to be explored in depth in the final stage of the review.

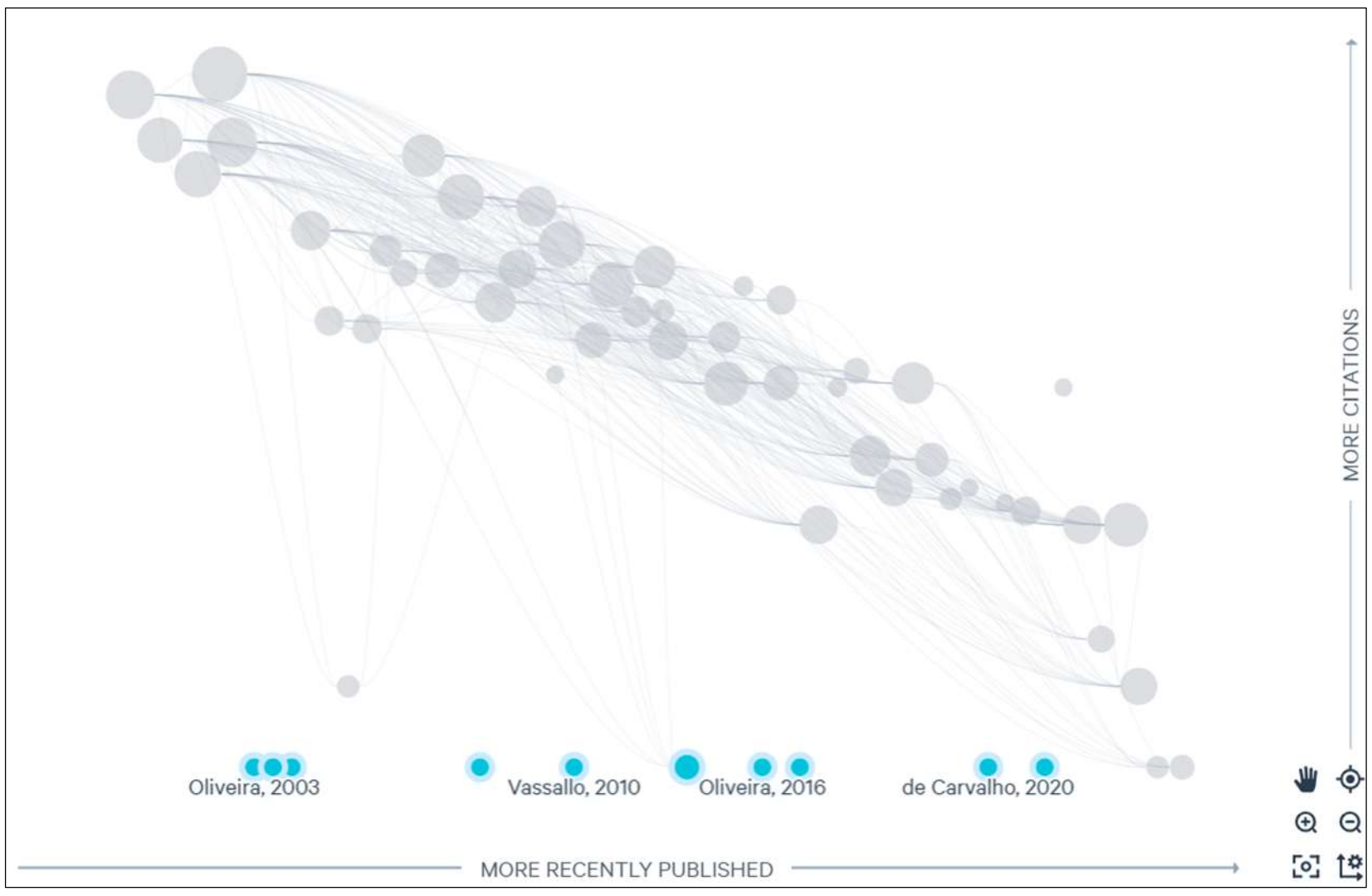

**Figure 9 - Articles researched and added to the final database, related to cooperation practices in the Brazilian context (Own elaboration. Source: Litmaps)**

## III.3. DATABASE OF BRAZILIAN STUDIES

Based on the final consolidated database, the analysis to search for the main articles on codeshare agreements in Brazil resulted in the identification of seven articles of interest for the objectives of this study, of which one was theoretical and six were empirical. These practical studies will therefore be the subject of in-depth analysis to ascertain the main evidence and contributions to the Brazilian scenario. Figure 10 below highlights these articles selected from the final review database, according to a map prepared in Litmaps.

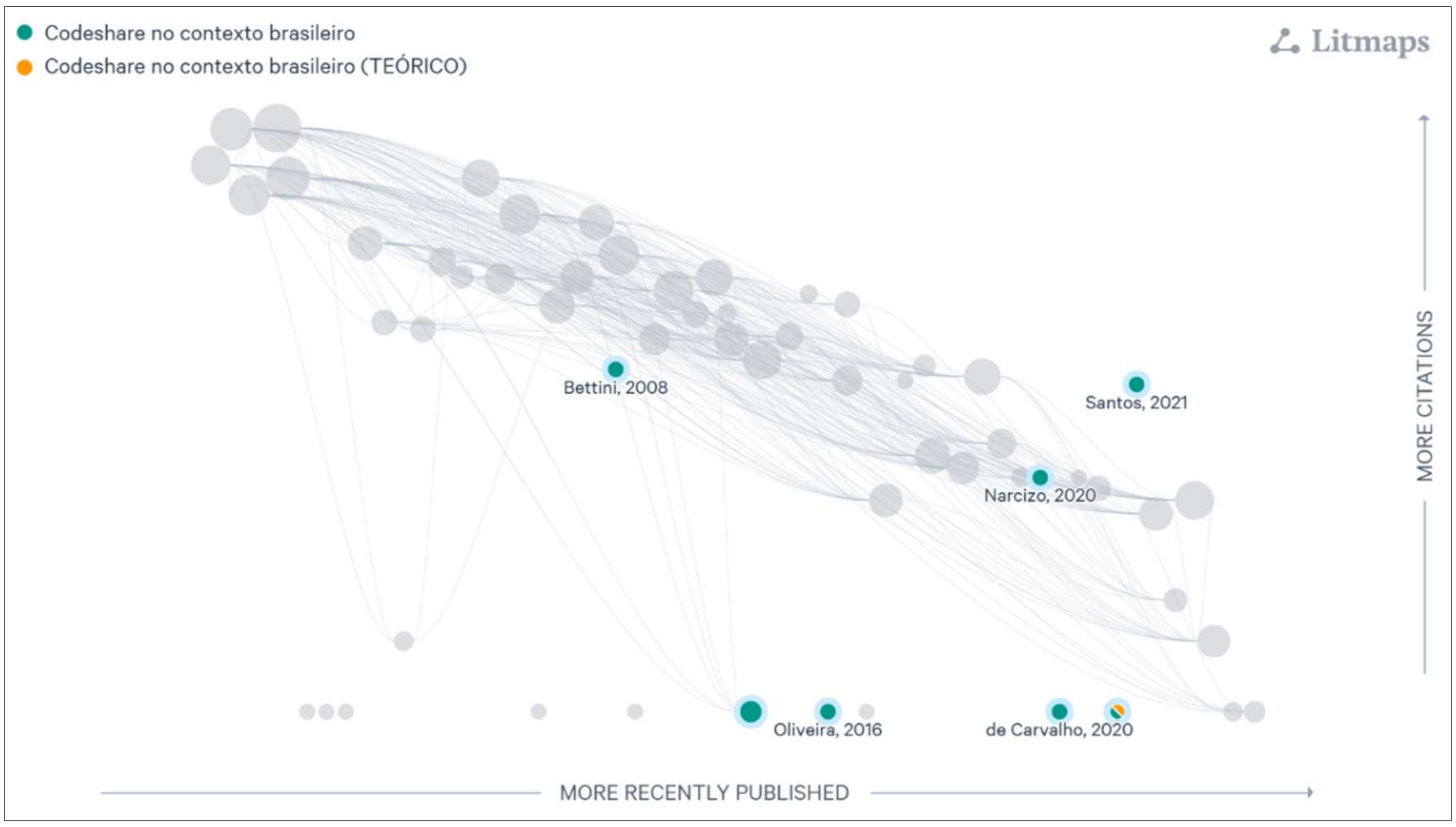

**Figure 10 - Main articles identified in the final database for further analysis according to the study objectives (Own elaboration. Source: Litmaps)**

Considering that the objective of this study is to identify and evaluate empirical evidence on the effects of codeshare agreements, the article by Britto (2021), although it offers a useful discussion on regulatory aspects and points to be observed in the analysis of these agreements in the Brazilian context, will not be used as a reference for the central conclusions of the review. This is because its contribution is predominantly conceptual and normative, without presenting empirical results that allow direct comparisons with the other studies analyzed. However, the article presents an important and interesting overview of the history of codeshare agreements signed in Brazil since 1998, drawing a parallel with the development and maturation of the antitrust regulatory body operating in the Brazilian market, the Administrative Council for Economic Defense (CADE), over the years.

Figure 11 below presents a timeline of the various codeshare agreements in the Brazilian context, on domestic and international routes, between Brazilian airlines or between Brazilian and foreign airlines since the agreement established between Vasp and Transbrasil in 1998, to the recent and now terminated codeshare agreement between Azul and Gol, which began in 2024.

Finally, the final overview of six selected empirical articles presents a snapshot of scientific production on empirical studies on codeshare agreements, considering Brazilian destinations or airlines, over the years. The research methodology resulted in empirical articles for this group of selected publications that evaluate these practices as the central focus of the study, deepening the analysis and discussions about codesharing itself. In addition, practical articles were also selected that evaluate these agreements under routes as a secondary factor in more comprehensive studies, which allows for the analysis of the influence of this tool in a manner that complements the other aspects studied.

The articles selected in this group were published from 2008, with the study by Bettini and Oliveira (2008), to the most recently published article by Santos et al. (2021), and represent examples of case studies that analyze agreements signed between large Brazilian airlines. In addition, most of the selected studies, five out of six, are characterized as econometric studies, while Narcizo et al. (2020) is the only empirical study that does not focus directly on the analysis of econometric aspects of the market, evaluating aspects of fleet diversification of Brazilian airlines in its publication.

| Period / Phase | Year(s) | Agreements / Details |
|---|---|---|
| First Codeshares in Brazil | late 1990s | VASP/Transbrasil |
| | early 2000s | Transbrasil/TAM |
| | | TAM T. A. Regionais S.A. / TAM T. A. Meridionais S.A. |
| | | Varig/TAM |
| Agreements prior to the "New SBDC Law" (2011) | late 2000s | TAM/TAP and Oceanair/BRA |
| | | TAM/United Airlines and TAM/LAN |
| | | Varig/Air France |
| | | TAM/US Airways and TAM/Continental |
| SBDC Regulatory Framework and Determinations on CADE's Role | early 2010s | Varig/Webjet |
| | | Varig/Iberia and Trip/Azul |
| | | TAM and international airlines (American Airlines, South African, Iberia, BA) |
| | | TAM and national or regional airlines (Varig, Trip, ADSA and Passaredo) |
| | | TAM/Iberia |
| | | TAM/AA and TAM/Passaredo (Voepass) |
| New resolutions from 2016 onwards for agreements shorter than 2 years | late 2010s on | Latam/Aeroméxico |
| | | Latam/Azul |
| | | Gol/Alitalia |
| | | Azul/GOL |

**Figure 11 - History of codeshare agreements signed in the Brazilian context, in parallel with the development of the regulatory framework over the years, from 1998 to 2025 (Own elaboration; source: Britto, 2021)**

Table 1 below presents the articles identified and selected for the in-depth review proposed as the final stage of this study, arranged with a summary of their main information, evidence obtained, and contributions to the topic.

**Table 1 - Main empirical articles on codeshare agreements in the Brazilian market identified by the study (Own elaboration; source: Britto, 2021)**

| Ref. Litmaps | Study | Period | Codeshare | Empirical Evidence | Analysis of Codeshare |
|---|---|---|---|---|---|
| Costa, 2015 | Price effects of codeshares vs. LCCs | 2001–2004 | Varig–TAM (2003–2005) | LCCs generated fare reductions, smaller on routes with codeshare | Central focus of the study |
| Bettini, 2008 | Airline seat capacity after reregulation | 1998–2004 | Varig–TAM (2003–2005) | Evidence of anti-competitive effects through reduced seat supply | Central focus of the study |
| Santos, 2021 | Impact of COVID-19 on air travel demand | 2010–2018 / 2019 | LATAM–Azul (2020–2021) | Codeshare had no significant effect in mitigating the demand drop during the pandemic | Secondary factor |
| Carvalho, 2020 | Effects of credit availability on air travel demand | 2002–2013 | Varig–TAM (2003–2005) | Positive coefficient for air travel demand | Secondary factor |
| Narcizo, 2020 | Fleet standardization and the impact of mergers | 2002–2017 | Varig–TAM (2003–2005) | Codeshare, among other factors, contributed to greater fleet diversification | Secondary factor |
| Oliveira, 2016 | Competitive impacts of regulatory slot-redistribution rules at airports | 2000–2005 | Varig–TAM (2003–2005) | Evidence of increased market power for cooperating airlines and others; Airports may be affected, regardless of being slot-constrained or not | Secondary factor |

Among the two articles identified that consider codeshare agreements as the central object of the study, the publication by Costa and Oliveira (2015) stands out for its assessment of the effects of these cooperation agreements on airfare prices, with the aim of evaluating the impact of airlines' positioning when cooperating to compete with competitors that operate as low-cost airlines. The article considered as a case study the codeshare agreement signed between Varig and TAM in the early 2000s, using data from operations on 86 Brazilian domestic routes between September 2001 and March 2004, and applying an econometric model with fixed effects. The results obtained by the author show that the practice of codesharing by airlines in the market, even if it does not cancel out the impacts of a new low-cost competitor on a given route, contributes to softening the effect on price reductions and, consequently, competitive pressures in the market: while routes without codesharing showed an average reduction of 12.36% with the entry of a new low-cost airline, the reduction caused on routes where codesharing agreements are already in place is lower, at around 8.30% on average, due to the degree of coordination between the actions and measures of the airlines in these markets.

The other article identified that presents codesharing as a central factor in its empirical study is the publication by Bettini and Oliveira (2008), in which the authors sought to identify the main elements influencing the decision-making of airlines operating regular flights in Brazil in relation to capacity offered during the important period of re-regulation of the Brazilian airline market. The study used data from 1998 to 2004 to evaluate seat supply indicators before and during the re-regulation period evaluated, and assessed the codeshare agreement signed between Varig and TAM, highlighting the methodology and scope defined. As main results, the author obtained empirical evidence that the regulatory measures taken during this period, including the codeshare agreement signed between Varig and TAM, had an influence in reducing seat supply in the market, as an anti-competitive effect of these cooperation practices.

In addition to these two articles, which highlight the analysis of codesharing, four other empirical studies were identified that consider codeshare agreements as secondary factors, associating the analysis of their effects with other phenomena or factors taken as a basis in the studies. Santos et al. (2021), for example, comprise the group of main articles identified by this review and included a secondary variable representative of the recent codeshare agreement signed between LATAM and Azul in Brazil in their empirical analysis of the impacts of the COVID-19 pandemic on flight demand in Brazil, based on data from the first year of the pandemic, 2019, in relation to the previous years. Among the various results and contributions of the author in the article, it was evident that the codeshare agreement evaluated by the study, which had been signed with the objective of circumventing high operating costs and market restrictions during the pandemic, did not prove to be significant in mitigating the drop in demand for air transport in Brazil during the period.

Another article that evaluates codeshare agreements in Brazil as a marginal factor among the main ones identified, Oliveira (2016) developed an econometric study on the impacts on competitiveness of regulatory measures for slot redistribution, applied in the Brazilian airline sector during the re-regulation period, starting in the 2000s. The author used a Relative Market Power (RMP) model to assess aspects of frequency concentration at congested airports in relation to the effective market power of cooperating airlines, using data from the early 20th century, and also evaluating the codeshare between Varig and TAM as a reference in the econometric analyses. The main results obtained indicate that slot redistribution actions in favor of airlines with smaller market shares would generate greater deconcentration of flights on congested routes and airports and, consequently, greater welfare for society. In addition, evidence from the study also indicates that airports may suffer consequences from the high concentration and market power of airlines, regardless of whether or not they are slotted, indicating the incidence of anti-competitive effects due to the higher degree of cooperation in the sector in these cases.

The remaining articles among the main ones selected also analyze codesharing in Brazil as a secondary factor in empirical modeling. The article Carvalho et al. (2020) develops an econometric study based on socioeconomic data and air transport demand between January 2002 and December 2013 to assess the influence of credit availability on air demand in Brazil. The author adds the codeshare agreement signed between Varig and TAM as a binary variable in the defined model, whose results were not statistically significant in the robustness tests performed, indicating the need for additional studies or variables to better assess the influence of codesharing in this context. The author concludes the article without making any relevant considerations or discussions about codesharing.

The article by Narcizo et al. (2020), the last among those identified for further study, presents a proposal for an empirical model to evaluate the standardization of Brazilian airline fleets, considering the dynamic impacts due to merger and acquisition procedures between Brazilian airlines. Using a database of monthly air traffic for the four major airlines at the time (TAM, Gol, Azul, Avianca) from January 2000 to December 2017, the "event study" proposed by the article analyzed several of these examples that occurred in the Brazilian airline market: from the merger between Varig and Gol in 2007 to the formation of Latam Airlines in 2012, adding the codeshare between Varig and TAM as a secondary variable in the proposed empirical model. The evidence obtained indicates that, more than the country's economic development or economic fluctuations, the occurrence of events such as airline bankruptcies, mergers/acquisitions, or cooperation through codesharing had a greater influence on the increase in the diversity of Brazilian airlines' aircraft fleets during the period.

The set of articles selected for the summarized scope of the proposed theme, considering practical studies on the practice of codesharing in the Brazilian airline market, allowed for the analysis of recent scientific contributions on the theme. The proposed methodology, using the Litmaps tool to optimize the research stages and visualization of review articles, resulted in a final database of articles on codesharing in Brazil and worldwide, and in a collection of articles identified on empirical analyses carried out in the Brazilian context, taken as a basis for the literature review and obtaining the results according to the proposed objective.

# IV. Patterns in Brazilian literature on codeshare agreements

The group of articles that is the main focus of this stage of the literature review consists of six empirical studies and one theoretical study. This set serves as a snapshot of the existing literature and shows that, despite the extensive international production on codesharing, the number of studies specifically focused on the Brazilian context is still small. The final review database, structured according to the established methodology, contains numerous studies on codesharing in different markets and situations. This volume contrasts with the set of studies focused on the Brazilian case identified at the end of the review, although this proportion is compatible with the participation of Brazilian academic production on air transport in the international literature.

When looking at the scope of codeshare agreements in Brazil, it is noted that most of the selected empirical studies focus on the analysis of the agreement signed between Varig and TAM between 2003 and 2005. This focus is natural not only because of the widespread attention received from the media and regulatory agencies, but also because the agreement occurred at a time of high concentration in Brazilian air transport, in which Varig and TAM held a dominant share of the market while other airlines, especially Gol, still had a reduced share. The only exception with s within the defined study group is Santos et al. (2021), which examines the agreement signed between Azul and Latam during the COVID-19 pandemic.

Analysis of the results presented in the selected articles reveals a predominance of evidence pointing to the negative effects of codeshare agreements on welfare in the Brazilian market. Among these effects, the increase in market power and the occurrence of anti-competitive impacts stand out, as shown by Bettini and Oliveira (2008), Costa and Oliveira (2015), and Oliveira (2016). There is also an example in which the effect of codesharing was not statistically significant as a secondary variable within the model, as in the case of Carvalho et al. (2020). In addition, it is observed that the agreement signed between Azul and Latam during the pandemic did not show relevant results in mitigating the drop in demand, according to Santos et al. (2021). Another point highlighted by the literature is the evidence that greater cooperation and the practice of codesharing contributed to an increase in the diversity of fleets operated by Brazilian airlines, as indicated by Narcizo et al. (2020).

The analyses carried out here suggest that the Brazilian literature on the subject is largely marked by the use of codesharing as a secondary variable, that is, present in empirical models, but not constituting the central research question. In these studies, codeshare appears as one of the explanatory elements within broader analyses, which makes its role relevant, but not treated as the main focus. This characteristic may induce artificial intelligence-based tools, such as Litmaps, to prioritize only articles whose central theme is explicitly codesharing, failing to recognize studies in which the practice has an important influence, even though it is not the direct object of the investigation. This potential bias reinforces the need for increased attention on the part of researchers. The use of AI increases productivity and facilitates bibliographic exploration, but it must always be accompanied by careful and judicious reading to avoid neglecting significant contributions that do not readily emerge in the results suggested by the algorithms.

# V. Conclusions

This study sought to review the literature on codeshare agreements, with an emphasis on the Brazilian context, using Litmaps as an artificial intelligence-based support tool for organizing and analyzing publications. The evidence indicates that the international literature presents ambiguous effects associated with these agreements, that is, mixed results, with some cases indicating benefits and others pointing to losses for society. However, we identified that the Brazilian literature on the subject shows a distinct pattern. In the studies evaluated, the estimated anti-competitive impacts appear more consistently than possible benefits, indicating a predominance of studies that point to increased market power and reduced competition in the situations analyzed. Thus, the discussion about the balance between operational gains for airlines and potential competitive losses is recurrent in the literature, and, in the Brazilian case, the results suggest that the effects linked to market concentration and strategic cooperation were more relevant than direct improvements for consumers or for the efficiency of the system.

When considering the set of scientific publications applied to Brazil, it is observed that there is still significant room for expanding empirical research on the topic. The studies identified focus on a few examples of agreements and fail to cover other relevant cases that have occurred over the years, including those involving foreign companies or international routes. The absence of analyses of these episodes represents a gap in the literature and points to research opportunities that could contribute to a more comprehensive understanding of the effects of codesharing in the Brazilian market.

A final point to highlight refers to how literature reviews are conducted in fields that use empirical methods and broad sets of explanatory variables, such as econometrics and machine learning. Many studies may present relevant results associated with variables that are not the main focus of the research but contribute to a more complete understanding of the phenomenon analyzed. In a context of continuous expansion of scientific production and increasing use of artificial intelligence tools for bibliographic organization, there is a possibility that these results will receive less attention, given that algorithms tend to prioritize works whose central theme explicitly coincides with the search terms. Thus, even with the support of these tools, it remains important for researchers to read carefully and interpret thoughtfully in order to prevent potentially useful contributions, even if secondary in the original studies, from being left out of literature reviews.

## Acknowledgments

The second author would like to thank the National Council for Scientific and Technological Development (CNPq), grant no. 305439/2021-9, and the São Paulo Research Foundation (FAPESP), grant no. 2024/01616-0. The authors thank Mauro Caetano, Marcelo Guterres, Evandro Silva, Giovanna Ronzani, Rogéria Arantes, Cláudio Jorge Alves, Bruno F. Oliveira, and the participants of the Air Transport Economics Symposium (SETA25). All errors and omissions are attributed to the authors.

**Communications in Airline Economics Research (CAER) is an open-access publication of academic articles from the Air Transport Economics Center of the Technological Institute of Aeronautics (ITA) (NECTAR-ITA). This series of studies focuses on analyzing the economic aspects of the air transport industry in Brazil. The works published in CAER are produced according to rigorous standards of scientificity and reproducibility, submitted to a careful internal evaluation, but do not undergo external peer review. The publication aims to be a platform for researchers and students in the field of Air Transport Economics, enabling them to present the relevance of their studies to society, contributing to the development of the Brazilian economy.**